\documentclass[letter]{aa}
\usepackage{graphicx}
\usepackage{txfonts}
\usepackage{natbib}
\usepackage{xcolor}

\usepackage{amsmath}
\usepackage{bm}
\usepackage{comment}
\newcommand{\Myr}{\mathrm{Myr}}

\begin{document}
\title{Stellar age is not disk age}
\subtitle{Late infall and the interpretation of disk fractions}
  \author{M. Kuffmeier
          \inst{1}
          \and
          A. Winter\inst{2}
          \and
          A. Kuznetsova\inst{3}
          \and
          M. Vioque\inst{4}
          \and
          A. Gupta\inst{5}
          }
\institute{
Niels Bohr Institute, University of Copenhagen, Jagtvej 155a, 2200 Copenhagen N, Denmark\\
\email{kueffmeier@nbi.ku.dk}
\and
{ School of Physical and Chemical Sciences, Queen Mary University of London, London, E1 4NS, UK}
\and
{Department of Physics, University of Connecticut, 196A Auditorium Road, Unit 3046, Storrs, CT06269, USA}
\and
{European Southern Observatory, Karl-Schwarzschild-Str. 2, 85748 Garching bei München, Germany}
\and
{Department of Astronomy, University of Virginia, Charlottesville, VA 22904, USA}
}
\date{Received \today}
\abstract
{Disk fractions in star-forming clusters decline with stellar age, and are routinely read as the lifetime distribution of primordial protoplanetary disks. We argue that this reading is not warranted, because late infall can replenish or reform disks after collapse. To formalize this, we model the disk fraction as a superposition of primordial and environmentally replenished components, with region-to-region dispersion in the supply rate. Disk fractions cannot distinguish a long-lived disk from a short-lived one that is repeatedly replenished. In Sco--Cen both reproduce the data, the replenished model marginally better. Stellar age therefore need not equal disk age, and stars can host successive disk generations with distinct mass, orientation, and chemistry. A key test is whether regions of the same stellar age show a genuine spread in disk fractions, as environmental variation in the replenishment supply predicts.}
\keywords{
protoplanetary disks --
stars: formation --
accretion, accretion disks --
ISM: clouds --
planets and satellites: formation
}
\maketitle
\section{Introduction}
\label{sec:intro}
Disk fractions as a function of stellar age are among the most widely used empirical constraints on the timescale available for planet formation.
In the canonical picture, a disk forms once during the collapse of a magnetized prestellar core \citep{Shu+1987,Tsukamoto+2023}, evolves through accretion, winds, dust growth, planet formation, and photoevaporation, and finally disperses \citep{Lissauer1993}. The age of the star is then taken to be approximately the age of the disk.
The fraction of stars with infrared or millimeter excess decreases from nearly unity in the youngest regions to small values after several million years \citep{Haisch+2001,Mamajek2009,Ribas+2015,Richert+2018,Polnitzky+2026}, and this decline is commonly interpreted as the dispersal of a single primordial disk population \citep[][and references therein]{Alexander+2014,Manara+2023}.
However, circumstellar disks do not form and evolve in isolation. Stars and their primordial disks are born in turbulent, filamentary molecular clouds \citep{Andre+2014,Hacar+2023}, and their interaction with the ambient medium need not cease when the initial collapse phase ends. Scattered-light observations reveal large-scale shadows and spiral arms around Class~II disks \citep{Garufi+2026}, and infalling streamers are now reported across all evolutionary stages \citep[see review by][]{Pineda+2023} -- in Class~0 protostars \citep{Pineda+2020,Valdivia-Mena+2024,Gieser+2025}, Class~I sources \citep{Yen+2019,Segura-Cox+2020,Valdivia-Mena+2022,Segura-Cox+2023,Flores+2023}, and Class~II disks \citep{Huang+2020,Huang+2021,Ginski+2021,Speedie+2025}. Statistics of reflection nebulae provide further evidence for widespread infall \citep{Gupta+2023}.

The idea that environmental accretion can reform disks has a theoretical history. \citet{Padoan+2005} and \citet{ThroopBally2008} showed that Bondi--Hoyle accretion after the initial collapse can exceed the primordial disk mass, which \citet{MoeckelThroop2009} confirmed with hydrodynamical simulations, and \citet{Scicluna+2014} argued analytically that a substantial fraction of pre-main-sequence stars capture enough material to form a second-generation disk. Motivated by these results and by simulations of star formation in turbulent clouds showing that stars accrete substantial post-collapse material \citep{Kuznetsova+2018}, including gas initially located beyond the prestellar core \citep{Padoan+2014,Kuffmeier+2017,Kuffmeier+2023,Pelkonen+2021}, \citet{Dullemond+2019} replaced the face-on geometry of the classical Bondi--Hoyle--Lyttleton (BHL) formulation with encounters at a finite impact parameter. Follow-up studies using this ``cloudlet capture'' setup demonstrated that late infall can form a second-generation disk \citep{KuffmeierGoicovicDullemond+2020} even around a surviving primordial one \citep{Kuffmeier+2021}, consistent with the misaligned disks found in population-synthesis calculations \citep{Bate2018}. \citet{Winter+2024} and \citet{Padoan+2025} further argued that late accretion in turbulent clouds contributes significantly to the mass and angular momentum in disks of the observed Class II population.

Observationally, there are several indications for multigenerational disks: accretion rates at ages where accretion should have ceased \citep{Delfini+2025}, ``Peter Pan'' disks around low-mass stars at ages $\gtrsim20$ Myr
\citep{Silverberg+2020}, widespread inner-disk misalignments seen in scattered light \citep{Villenave+2024}, and the reported correlation between spatial location and accretion rate in Lupus \citep{Winter+2024}.
Moreover, the distribution of stellar obliquities \citep[see the review by][and references therein]{Albrecht+2022}, the frequent misalignments of planet-forming disks \citep{Biddle+2025} and debris disks \citep{HurtMacGregor2023}, and high-eccentricity debris disks \citep{Lovell+2026} further indicate infall.
Considering that late infall replenishes disks, or even forms new ones after the primordial disk has dispersed, the observed disk fraction is not the survival probability of a single disk generation but a superposition of primordial and rejuvenated disks. In this Letter we formalize this distinction with a simple population model consisting of a dispersing primordial disk and replenishment that can explain the observed disk fractions. We discuss observational signatures of disks that are replenished by infall.

\section{A population model for multigenerational disks}
\label{sec:model}
\begin{figure}
    \centering
    \includegraphics[width=0.85\linewidth,trim={15 5 5 0cm},clip]{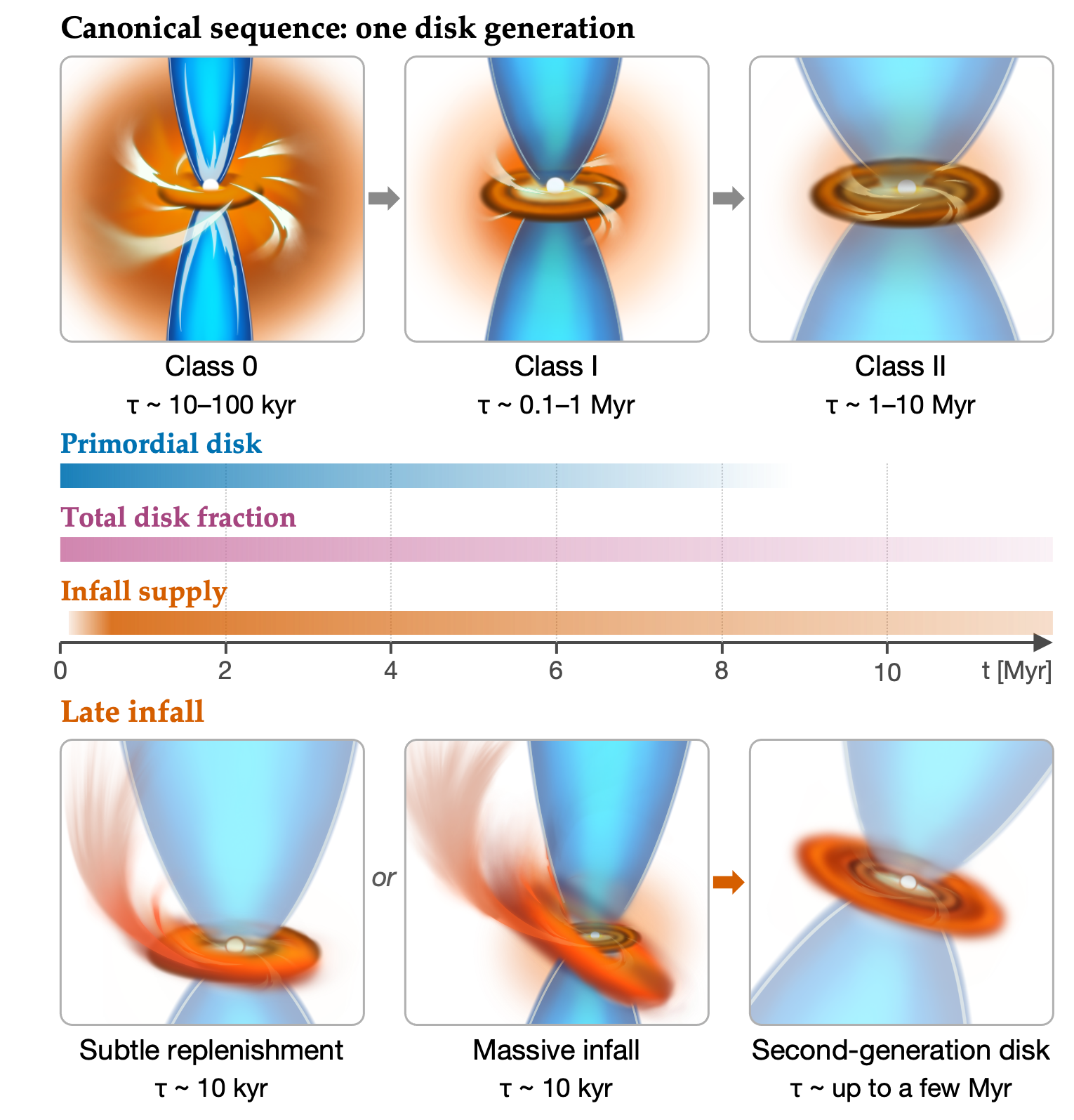}
    \caption{\textnormal{Schematic of the proposed replenishment framework. Top: Standard sequence from an embedded protostar (Class 0) to an embedded disk with outflow (Class I) to a protoplanetary disk (Class II), leaving a primordial disk. Bottom: Short late infall events replenish a disk or induce a new one, raising the disk fraction at later ages.}}
    \label{fig:sketch}
\end{figure}

The two pathways we consider are shown schematically in Fig.~\ref{fig:sketch}: the canonical Class~0~$\rightarrow$~Class~I~$\rightarrow$~Class~II evolution along the top row, and the post-collapse infall pathway along the bottom, in which environmental accretion delivers fresh material that can replenish or even reform a disk.
The disk when observed at a stellar age, $t_\star$, is not necessarily the remnant disk that formed during the initial collapse. We give every disk generation a single intrinsic dispersal timescale and let the observed disk fraction emerge from the interplay between that dispersal and a time-dependent, stochastic supply of replenishment events. The free population-level parameters are introduced below.
\paragraph{Intrinsic disk dispersal.}
Every disk, whether inherited from the collapse phase or built later from infall, is assumed to have a constant differential dispersal probability, $\lambda_{\rm disk}\,{\rm d}t$, so that a disk present at $t=0$ survives to stellar age $t_\star$ with probability
\begin{equation}
    P_{\rm prim}(t_\star)
    = \exp\left(-\lambda_{\rm disk}\, t_\star\right)
    = \exp\left(-t_\star/\tau_{\rm disk}\right),
    \qquad
    \tau_{\rm disk}=\lambda_{\rm disk}^{-1}.
    \label{eq:pprim}
\end{equation}
This recovers the familiar exponential disk-fraction law, with the crucial reinterpretation that $\tau_{\rm disk}$ is the survival time of a single disk generation, primordial or replenished.

\paragraph{Replenishment as a nonhomogeneous Poisson process.}
Replenishment events are modeled as a nonhomogeneous Poisson process whose rate decays as the reservoir of dense environmental gas disperses,
\begin{equation}
    \lambda_{\rm supply}(t_\star \,|\, \lambda_0)
    = \lambda_0 \, \exp\left(-t_\star/\tau_{\rm rep}\right),
    \label{eq:lambdasupply}
\end{equation}
where $\lambda_0$ is the initial event rate and $\tau_{\rm rep}$ the population-level decay timescale of the supply, i.e.,\ the timescale over which the environment ceases to be able to feed disks.
A replenishment event delivers enough material to sustain or even reform a disk. 
For instance, infall providing a disk mass of $M_{\rm disk}\sim 10^{-2} M_{\star}$ is enough to produce a near-infrared-excess disk that then disperses. For simplicity, we assume that the dispersal happens at the same rate as a primordial one. 
The mean number of surviving replenished disks at age $t_\star$, $n(t_\star \,|\, \lambda_0)$, is the convolution of supply and survival and has a simple closed form (Appendix~\ref{app:derivation}, Eq.~\ref{eq:Mapp}). The probability of hosting at least one replenished disk follows from Poisson statistics,
\begin{equation}
    P_{\rm rep}(t_\star \,|\, \lambda_0)
    = 1 - \exp\left[-n(t_\star \,|\, \lambda_0)\right].
    \label{eq:prep}
\end{equation}
A star hosts a detectable disk if its primordial disk survives or if at least one replenished disk is present. Treating the two as independent at fixed $\lambda_0$, the disk fraction is the superposition
\begin{equation}
    P_{\rm disk}(t_\star \,|\, \lambda_0)
    = P_{\rm prim}
    + P_{\rm rep}
    - P_{\rm prim}\,P_{\rm rep},
    \label{eq:pdisc_fixedlam}
\end{equation}
where the final term avoids double-counting. Equation~(\ref{eq:pdisc_fixedlam}) is central: the disk fraction is a disk-hosting probability, set by the balance of dispersal and replenishment, rather than the survival curve of one disk.
Processes such as viscous evolution \citep{Lynden-BellPringle1974}, magnetohydrodynamic winds \citep{Lesur+2023}, multiplicity \citep{Cuello+2023}, external photoevaporation \citep{WinterHaworth2022}, and dynamical encounters \citep{Pfalzner+2005} remain important for the evolution of individual disks. Our claim is that environmental replenishment additionally affects disk demographics to a degree that has not yet been quantified.

\paragraph{Region-to-region dispersion in the supply rate.}
The initial replenishment rate, $\lambda_0$, encodes the local ambient gas density, the stellar number density, and the cross section for BHL-like capture (Appendix~\ref{app:bhl}). It is therefore not a universal constant but varies from star to star and, more importantly, from region to region. We model this variation with a lognormal distribution, $\ln\lambda_0 \sim \mathcal{N}(\mu_{\ln\lambda},\,\sigma_{\ln\lambda}^2)$, parameterized by its arithmetic mean, $\overline{\lambda}_0$, and log-dispersion, $\sigma_{\ln\lambda}$ (Appendix~\ref{app:derivation}). 
\textnormal{The initial rate,}
\begin{equation}
    p(\lambda_0 \,|\, \bm{\theta})
    = \frac{1}{\lambda_0\,\sigma_{\ln\lambda}\sqrt{2\pi}}
      \exp\!\left[
      -\frac{(\ln\lambda_0-\mu_{\ln\lambda})^2}{2\sigma_{\ln\lambda}^2}
      \right],
    \label{eq:lognormaldensity}
\end{equation}
is parameterized so that $\overline{\lambda}_0$ is the mean of the distribution, which fixes the log-space location at
$\mu_{\ln\lambda}=\ln\overline{\lambda}_0-\tfrac{1}{2}\sigma_{\ln\lambda}^2.$
A single lognormal thus absorbs both the star-to-star and the region-to-region variation in the supply rate. Physically, $\lambda_0$ depends on the ambient density and velocity field
(Appendix~\ref{app:bhl}), which vary far more between regions than within them,
so we interpret $\sigma_{\ln\lambda}$ as primarily region-to-region variation.
Marginalizing Eq.~(\ref{eq:pdisc_fixedlam}) over $\lambda_0$, and since $P_{\rm prim}$ is independent of $\lambda_0$, the population-averaged disk fraction is
\begin{equation}
    f_{\rm disk}(t_\star \,|\, \bm{\theta})
    = P_{\rm prim}(t_\star)
    + \bigl[1 - P_{\rm prim}(t_\star)\bigr]\,
      f_{\rm rep}(t_\star \,|\, \bm{\theta}),
    \label{eq:fdisc_population}
\end{equation}
with the population-averaged replenished fraction
\begin{equation}
    f_{\rm rep}(t_\star \,|\, \bm{\theta})
    = \int P_{\rm rep}(t_\star \,|\, \lambda_0)\,
      p(\lambda_0 \,|\, \bm{\theta})\,{\rm d}\lambda_0 ,
    \label{eq:frep_population}
\end{equation}
evaluated numerically (Appendix~\ref{app:derivation}).

The dispersion, $\sigma_{\ln\lambda}$, provides a physically motivated origin for the documented scatter in disk fractions between regions of nominally the same age \citep{Polnitzky+2026}, and offers an explanation for the broad spread in disk-lifetime distributions. 
A region that has retained more dense gas (large $\lambda_0$) sustains a higher replenished fraction at evolved ages than a coeval region that dispersed its gas early (small $\lambda_0$), even when $\tau_{\rm disk}$, $\overline{\lambda}_0$, and $\tau_{\rm rep}$ are identical. 
The shorter apparent disk lifetimes for stars of higher masses ($1$--$3\,M_\odot$) \citep{Pfalzner+2026} are not in tension with infall, and plausibly support it. Because prolonged late infall is predicted to contribute a larger mass fraction to more massive stars \citep{Pelkonen+2021,Kuffmeier+2023}, their stellar age underestimates the time since they first hosted a disk, shortening the apparent lifetime.
\section{Constraints from disk fractions}
\label{sec:degeneracy}
\begin{figure}
    \centering
    \includegraphics[width=\columnwidth]{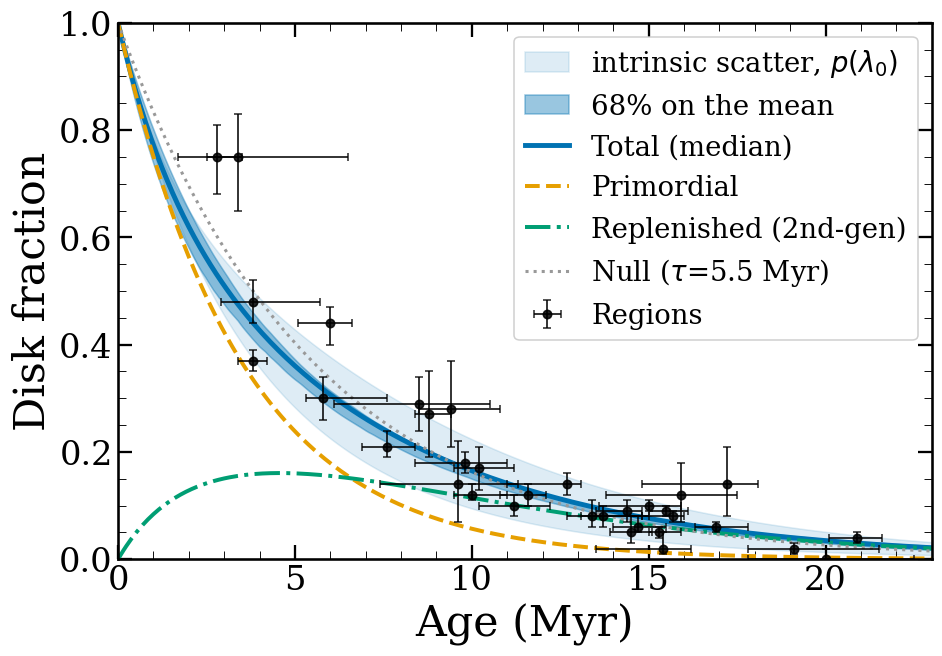}
    \caption{Decomposition of the disk fraction into primordial (Eq.~\ref{eq:pprim}, dashed) and replenished (Eq.~\ref{eq:frep_population}, dash-dotted) components, from the population fit of the multigenerational model to the Sco--Cen disk fractions of \citet{Polnitzky+2026} (black points and error bars). The dark band is the $68\%$ credible interval on the mean relation; the light band is the intrinsic region-to-region scatter induced by $p(\lambda_0)$. The dotted gray curve is a single-population model with $\tau=5.5\,\mathrm{Myr}$. }
    \label{fig:bayesfit}
\end{figure}
Figure~\ref{fig:bayesfit} shows the model fit to the Sco--Cen disk fractions
of \citet{Polnitzky+2026}. 
Sampling the posterior by Markov chain Monte Carlo yields an intrinsic dispersal timescale of
$\tau_{\rm disk}=3.6^{+0.6}_{-0.8}\,\Myr$, a mean initial supply rate of
$\overline{\lambda}_0=0.10^{+0.18}_{-0.06}\,\mathrm{Myr^{-1}}$, a supply decay
timescale of $\tau_{\rm rep}=6.4^{+2.6}_{-1.8}\,\Myr$, and a broad region-to-region
scatter of $\sigma_{\ln\lambda}=0.74^{+1.24}_{-0.47}$. The apparent decay time of the
combined curve is $\approx5.5\,\Myr$, substantially longer than $\tau_{\rm disk}$.

Both models reproduce the Sco--Cen disk fractions. 
A single long-lived
population with $\tau=5.5\,\Myr$ fits the data, as does the replenished model,
which is marginally preferred by the Bayesian information criterion
\citep[BIC;][]{Schwarz1978}, $\Delta\mathrm{BIC}\simeq3$. We do not rest our
case on this preference. For only $\sim\!30$ regions is the BIC a crude
approximation to the Bayes factor, and it accounts for neither the adopted
priors nor the parameter degeneracies (Appendix~\ref{app:corner}); two of the
four posteriors ($\overline{\lambda}_0$ and $\sigma_{\ln\lambda}$) are weakly
constrained and demonstrate viability rather than measurement. Part of the
region-to-region scatter may also stem from \textnormal{other} effects: heterogeneous stellar masses, varying binary fractions, contamination,
differing initial conditions, and isochrone-fitting errors.
(Fig.~\ref{fig:prim_sec} highlights the relative contribution of the primordial disk and replenishment to the disk fraction.) 

The substantive point is not that the data prefer replenishment, but that they
cannot exclude it, while replenishment is independently expected. Streamers and
late-infall signatures are now seen across all evolutionary stages, including
Class~II disks (Sect.~\ref{sec:intro}), so a replenished contribution to the
disk fraction is expected. The single-population reading
fits the same data only by assuming that this ongoing infall leaves the
demographics untouched, which the growing catalog of streamers renders
unlikely. What the data do bound is the observed decay time, which is an upper
limit on the survival time of an individual disk; the canonical value of several
million years should be read as such. Breaking the degeneracy requires the diagnostics of
Sect.~\ref{sec:predictions}, together with multi-scale models connecting
molecular-cloud dynamics to disk scales over million-year timescales.

\paragraph{Plausibility of the replenishment rate.}
The decomposition requires that a substantial fraction of the disks observed at intermediate ages be replenished rather than primordial. With $\tau_{\rm disk}=3.6\,\mathrm{Myr}$, the surviving primordial component at $t_\star=5\,\mathrm{Myr}$ is $\approx25\%$, against an observed Sco--Cen disk fraction of $\approx35\%$ \citep{Polnitzky+2026}; from Eq.~(\ref{eq:fdisc_population}) this implies $f_{\rm rep}\approx14\%$, i.e.,\ roughly one star in eight hosts a replenished disk, and about $30\%$ of the disk-bearing population is second-generation. 
\textnormal{Observations support these numbers.
Ambient scattered light is seen for $>20 \%$ of the youngest sources in near infrared observations \citep{Garufi+2026}, and a comparable fraction shows millimeter infall signatures with rates of $\gtrsim 10^{-8} M_{\odot}$ yr$^{-1}$ \citep{Gupta_DECO}.
Furthermore, disk orientation records the latest infall event from the large-scale molecular cloud \citep{Wijnen+2017c}, and therefore explains preferential alignment of disks in multiples \citep{Hsieh+2026} and spatial correlation of accretion to the molecular cloud environment \citep{Pittman+2025}.}

\section{Distinguishing replenished from surviving disks}
\label{sec:predictions}
\textnormal{Dust traps may prevent fast disk dispersal \citep[e.g.,][]{vanderMarel+2018}.
However, dust traps are frequent remnants of infall events \citep{Kuznetsova+2022,ZhaoLau+2026}, too.
The issue is that disk} fractions alone cannot distinguish between longer-lived primordial disks and shorter-lived disks experiencing substantial replenishment. 
However, the underlying processes
can be constrained through other observables.
Infall refuels the disks while viscous spreading, winds, and photoevaporation remove material.
If infall dominates over removal effects, such as photoevaporation,
a disk can be replenished \citep{Ooyama+2026}, and it carries the angular momentum of the captured material, which is
not correlated with the star or any earlier disk.
Inner--outer misalignments, warps, and counter-rotating gas should therefore be
more common among evolved disks in regions providing significant replenishment.

In case of replenishment, accretion should persist past the age at which
a single evolving population would have shut off, plausibly connected to the
low-accreting, spatially dispersed young stellar objects
\citep{Delfini+2025}.
Material accreted late need not
share the composition of the natal core \citep{vanDishoeck+2023}, which is why second-generation
disks should also carry volatile and isotopic signatures decoupled from the host
star, linking to the heterogeneity recorded in primitive solar-system materials
\citep{Larsen+2020,vanKooten+2026}, \textnormal{as recently invoked to explain the asymmetric, carbon-rich chemistry of HD~142527 \citep{Temmink+2026}}.

\section{Conclusions}
We argue that disk fractions as a function of stellar age should not be read directly as the lifetime distribution of single primordial disks. Stars continue to accrete environmental material after the collapse phase, replenishing existing disks or forming new ones through late infall.
In this framework a star may host multiple disk generations, each with a different mass, angular momentum, orientation, and composition, and the planet-formation clock may be reset.
Against the background of the growing number of streamers
associated with star--disk systems, late infall is likely frequent enough to
leave an imprint on the observed disk fractions rather than only affecting individual objects.
A decisive question is whether regions of the same stellar age show a genuine spread in disk fractions as
environmental variation in the replenishment supply predicts.
Resolving this requires more accurate
stellar ages. Disk-based dynamical masses, such as those recently derived for
Upper Scorpius \citep{Zallio+2026}, offer one promising route.
Future demographic studies should account for late infall alongside intrinsic disk survival.

\begin{acknowledgements}
We specifically thank Martine Lützen for her artwork in producing the six illustrations visualized in Figure 1. 
This research was supported in part by the Munich Institute for Astro-, Particle and BioPhysics (MIAPbP), which is funded by the Deutsche Forschungsgemeinschaft (DFG, German Research Foundation) under Germany's Excellence Strategy---EXC-2094---390783311. MK acknowledges funding by the Independent Research Fund Denmark (DFF Sapere Aude Grant: 5251-00016B). AG acknowledges support from NSF AST-2407547 and the David and Lucile Packard Foundation and the Virginia Institute of Theoretical Astronomy (VITA).
\end{acknowledgements}
\bibliographystyle{aa}
\bibliography{general.bib}
\begin{appendix}
\section{Closed-form supply integral and population averaging}
\label{app:derivation}
Every disk disperses with the constant rate $\lambda_{\rm disk}=1/\tau_{\rm disk}$, so a disk supplied at time $t'$ survives to age $t_\star$ with probability
\begin{equation}
    S(t_\star,t') = \exp\left[-\lambda_{\rm disk}(t_\star-t')\right].
    \label{eq:Ssurv}
\end{equation}
With the exponential supply of Eq.~(\ref{eq:lambdasupply}), the mean number of surviving replenished disks is the convolution of supply and survival, which -- being linear in $\lambda_0$ -- evaluates in closed form,
\begin{align}
    n(t_\star \,|\, \lambda_0)
    &= \int_0^{t_\star}
       \lambda_0\,\exp(-t'/\tau_{\rm rep})\,
       \exp[-\lambda_{\rm disk}(t_\star-t')]\,{\rm d}t' \nonumber \\
    &= \lambda_0\,
       \frac{
       \exp(-t_\star/\tau_{\rm rep})
       -
       \exp(-\lambda_{\rm disk}\,t_\star)
       }{
       \lambda_{\rm disk} - 1/\tau_{\rm rep}
       },
    \label{eq:Mapp}
\end{align}
valid for $\lambda_{\rm disk}\ne 1/\tau_{\rm rep}$. In the degenerate case $\lambda_{\rm disk}=1/\tau_{\rm rep}$,
\begin{equation}
    n(t_\star \,|\, \lambda_0)
    = \lambda_0\, t_\star \exp(-\lambda_{\rm disk}\,t_\star).
    \label{eq:Mlimit}
\end{equation}
The probability that at least one replenished disk is present is $P_{\rm rep}=1-\exp(-n)$ (Eq.~\ref{eq:prep}); expanding to first order in $n$ recovers the low-occupancy limit $P_{\rm rep}\simeq n$, in which $P_{\rm rep}$ scales linearly with $\lambda_0$.
Because $P_{\rm rep}$ is a nonlinear (saturating) function of $n\propto\lambda_0$, the population-averaged replenished fraction $f_{\rm rep}$ (Eq.~\ref{eq:frep_population}) has no elementary closed form and is evaluated numerically. 
The total disk fraction then follows from Eq.~(\ref{eq:fdisc_population}). In the low-occupancy limit $f_{\rm rep}\simeq\overline{\lambda}_0\,\langle n/\lambda_0\rangle$ is linear in $\overline{\lambda}_0$ and independent of $\sigma_{\ln\lambda}$; the dispersion manifests only once saturation ($n\gtrsim1$) becomes important for the high-$\lambda_0$ tail, which is where the region-to-region scatter in $f_{\rm disk}$ originates. This is also why $\sigma_{\ln\lambda}$ is only weakly constrained by the present data: the posterior is informative only through the fraction of regions that reach saturation.
\section{Bondi--Hoyle--Lyttleton capture and the supply rate}
\label{app:bhl}
The supply rate $\lambda_{\rm supply}$ is a demographic quantity that depends on the probability that a star encounters gas dense enough, slow enough, and with suitable angular momentum to build a disk. For a star of mass $M_\star$ moving through gas of density $\rho$ with relative velocity $v_{\rm rel}$, the characteristic BHL accretion rate is
\begin{equation}
    \dot{M}_{\rm BHL} = 4 \pi \frac{G^2 M_\star^2 \rho}{\left(v_{\rm rel}^2 + c_s^2\right)^{3/2}},
    \label{eq:bhl}
\end{equation}
where $c_s$ is the sound speed \citep{Hoyle1960}. In a turbulent cloud, $\rho$, $v_{\rm rel}$, and the angular momentum of the captured gas vary strongly in space and time, so replenishment is expected to be episodic -- in agreement with luminosity bursts \citep{Fischer+2023} and with infall rates from GMC simulations \citep{Padoan+2025,Pelkonen+2021,Kuffmeier+2023,Kaalva+2026}. The strong dependence of Eq.~(\ref{eq:bhl}) on the local density and velocity field is what makes $\lambda_0$ vary by orders of magnitude between environments, motivating the broad lognormal prior.
A disk-building event requires not only mass capture but also sufficient specific angular momentum $j_{\rm acc}$, giving a circularization radius $R_{\rm circ}=j_{\rm acc}^2/(G M_\star)$. Material with $R_{\rm circ}$ comparable to or larger than the existing disk radius can substantially expand, warp, or reform the disk \citep{KuffmeierGoicovicDullemond+2020}. Since $j_{\rm acc}$ reflects the local turbulent environment, the angular-momentum vector of the replenished disk need not align with that of the primordial disk \citep{Bate+2010,Bate2018,Kuznetsova+2019,Kuffmeier+2024}, naturally producing multigenerational disks with different orientations, mass reservoirs, and chemical histories.

\section{Posterior parameter correlations}
\label{app:corner}
\begin{figure}
  \centering
  \includegraphics[width=\linewidth]{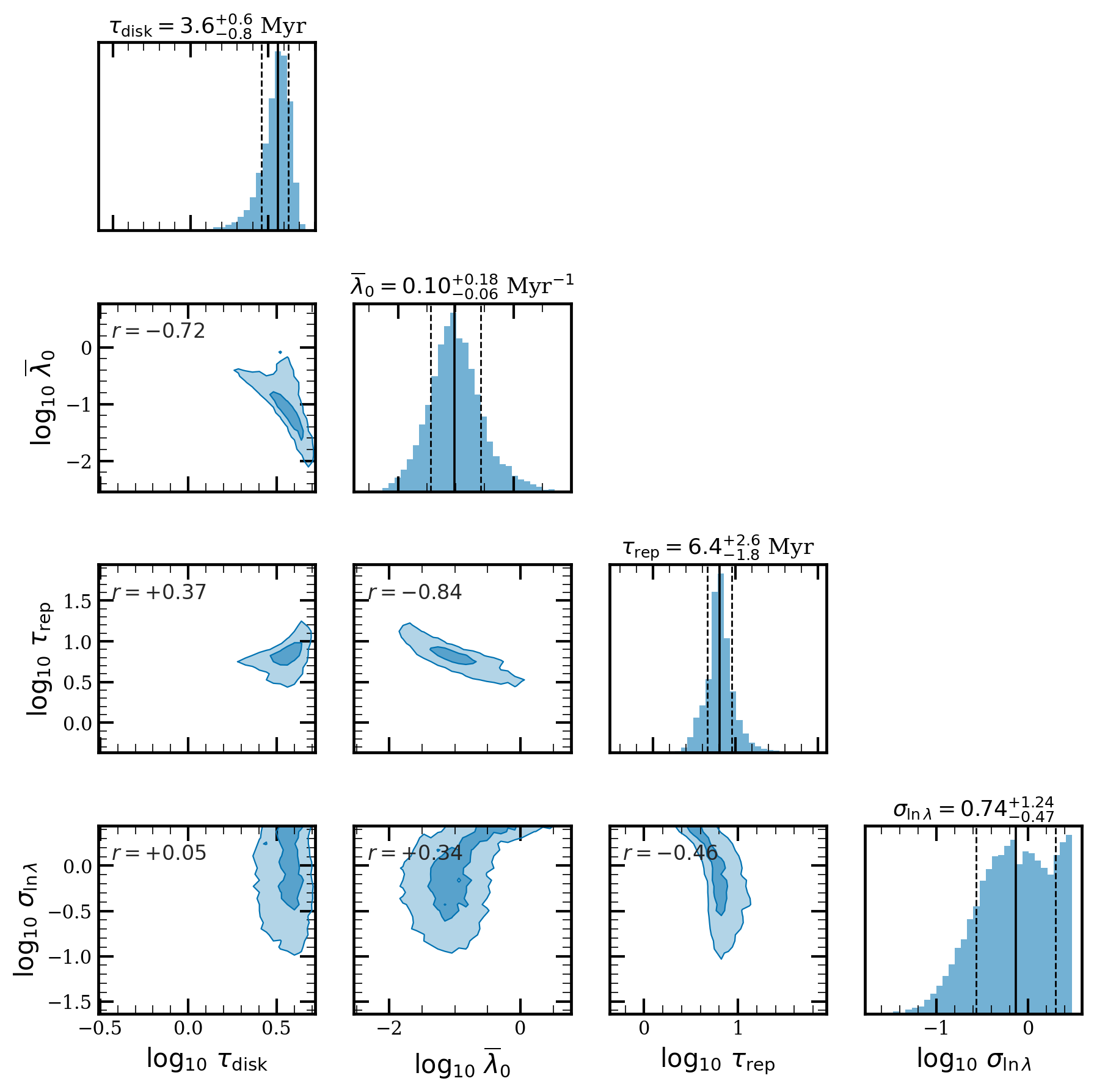}
  \caption{Joint posterior distribution of the four model parameters, shown on a
  base-10 logarithmic scale (the sampling space). Off-diagonal panels show the
  $1\sigma$ and $2\sigma$ joint credible regions, with the Pearson correlation
  coefficient (in $\ln$-space) annotated in each; diagonal panels show the
  marginal distributions with the median (solid) and 16th/84th percentiles
  (dashed), repeated as the median and 68\% interval
  above each panel. The two strongest degeneracies, $\overline{\lambda}_0$--$\tau_{\rm rep}$
  and $\tau_{\rm disk}$--$\overline{\lambda}_0$, reflect that the data constrain
  the time-integrated surviving supply rather than the individual parameters.}
  \label{fig:corner}
\end{figure}

The joint posterior of the four population parameters
$\bm{\theta}=\{\tau_{\rm disk},\overline{\lambda}_0,\tau_{\rm rep},\sigma_{\ln\lambda}\}$,
sampled as in Sect.~\ref{sec:degeneracy}, is shown in Fig.~\ref{fig:corner}.
The parameters are significantly correlated. The strongest degeneracy is between
the mean initial supply rate and its decay timescale,
$r(\ln\overline{\lambda}_0,\ln\tau_{\rm rep})\approx-0.84$: because the disk
fractions constrain chiefly the time-integrated, surviving replenishment
[Eq.~(\ref{eq:Mapp})], a higher initial rate can be traded against a shorter
supply duration. The intrinsic dispersal and supply rates are similarly
anti-correlated, $r(\ln\tau_{\rm disk},\ln\overline{\lambda}_0)\approx-0.72$,
since a shorter primordial lifetime is compensated by stronger replenishment.
The remaining pairs are weakly to moderately correlated ($|r|\lesssim0.5$).

These degeneracies are the parameter-space counterpart of the model comparison
in Sect.~\ref{sec:degeneracy}. They broaden the marginal posteriors -- most
strongly for $\overline{\lambda}_0$ and $\sigma_{\ln\lambda}$, which become
informative only through the high-$\lambda_0$ regions that reach saturation
(Appendix~\ref{app:derivation}) -- so that the individual parameters demonstrate
viability rather than constituting measurements. The directly observable
combination, the apparent decay time of the combined disk fraction
($\approx5.5\,\Myr$), is nonetheless well constrained and robustly exceeds
$\tau_{\rm disk}$, consistent with the central argument of this Letter.

\section{Primordial disk, replenishment and second-generation disk}
\label{app:prim_sec}
\begin{figure}
  \centering
  \includegraphics[width=\linewidth]{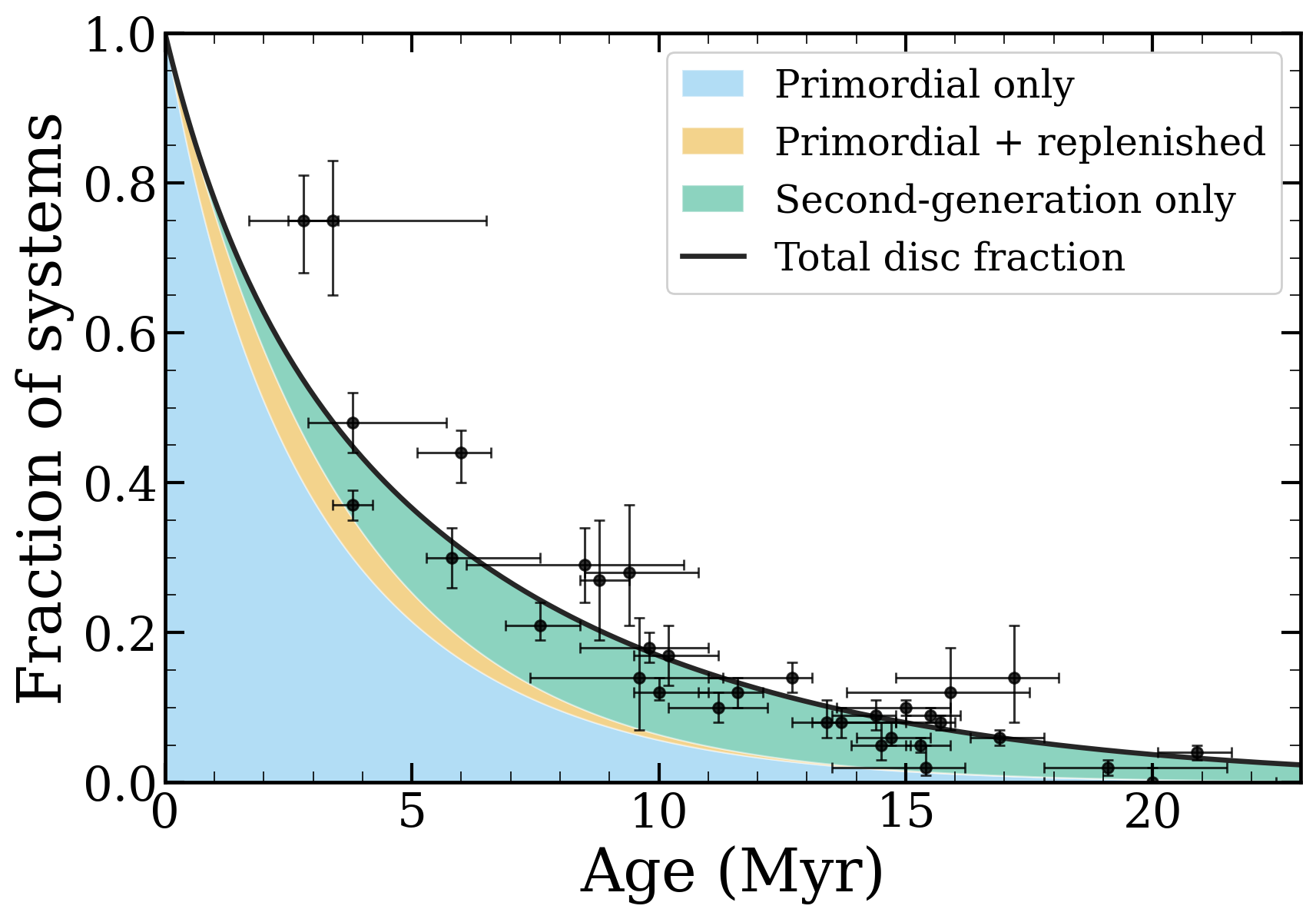}
  \caption{Same as Fig.~\ref{fig:bayesfit}, but only highlighting the relative contributions of primordial disk, primordial disk with replenishment and second-generation disks to the disk fraction based on the model.}
  \label{fig:prim_sec}
\end{figure}

\begin{figure}
  \centering
  \includegraphics[width=\linewidth]{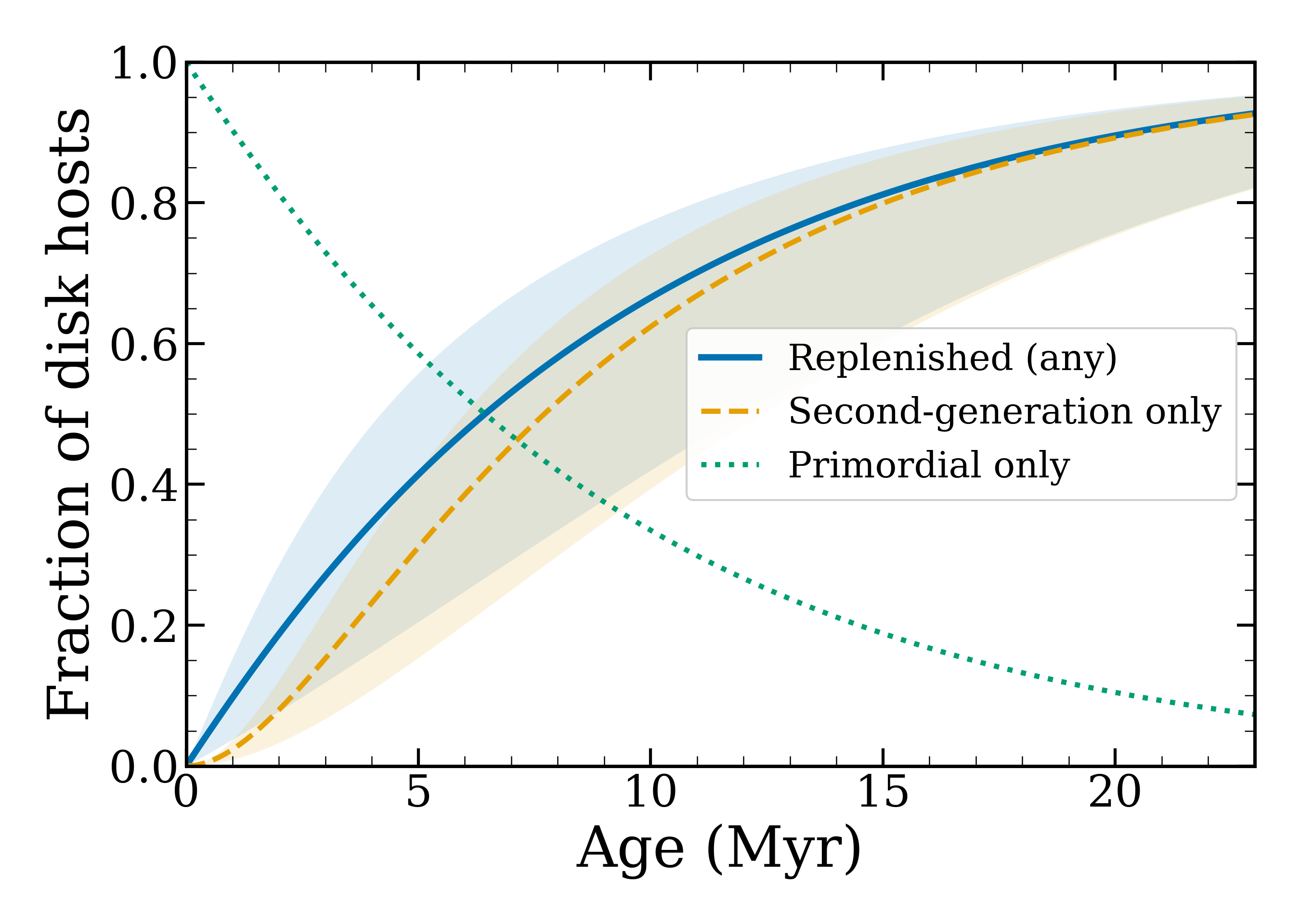}
  \caption{\textnormal{Composition of the disk-bearing population as a function of
  stellar age for the posterior-median model of Fig.~\ref{fig:bayesfit}: disks
  with any replenished component (solid), purely second-generation disks
  (dashed), and primordial disks that were never replenished (dotted). Bands
  give the region-to-region spread implied by $\sigma_{\ln\lambda}$.}}
  \label{fig:frac_repl}
\end{figure}

For illustration purposes of the relative contributions of replenishment and second-generation disks to the disk fractions at various ages, we show the model fit of Fig.~\ref{fig:bayesfit} without the uncertainties and color the individual contributions as primordial, primordial plus replenishment and second-generation disk (Fig.~\ref{fig:prim_sec}).
\textnormal{Figure~\ref{fig:prim_sec} shows disk fractions of all stars. High-resolution
imaging, chemical, and misalignment samples are instead selected on infrared
excess, so the relevant quantity is the composition of the disk-bearing
population alone. Figure~\ref{fig:frac_repl} shows the same posterior-median
model normalized by the total disk fraction, i.e.,\ conditioned on hosting a
disk: the share of disk hosts whose disk carries a replenished component, the
share whose disk is purely second-generation (the primordial disk has
dispersed), and the complementary share that is still primordial and never
replenished. Bands give the region-to-region spread implied by
$\sigma_{\ln\lambda}$ ($16$th--$84$th percentile of $\lambda_0$).}

\textnormal{At $1$--$3\,\Myr$ the model attributes $\approx10$--$27\%$ of the disk hosts to
any replenishment, and only $\approx2$--$15\%$ to a purely second-generation
disk: the typical young Class~II disk in Sco--Cen is expected to be primordial,
and the replenished population becomes the majority only beyond
$\approx7\,\Myr$. This is a prediction of the model rather than a difficulty
for it, and it is what distinguishes replenishment from an internal process
that stochastically prolongs disk lifetimes (Sect.~\ref{sec:predictions}): the
incidence of misalignments, streamers, and decoupled chemistry among disk hosts
should grow with stellar age, whereas a fixed internal channel predicts a
roughly age-independent incidence. It also implies that samples of
$\gtrsim5\,\Myr$ disk-bearing stars are the most efficient way to test the
framework.}

\textnormal{However, we emphasize that episodic, filamentary infall via streamers is already a prevalent feature during the early formation phase \citep{Kuffmeier+2017,Holm+2026}.
This implies that in reality the transition between primordial and replenished disk is more dynamic and less distinguishable than our simple model assumption suggests. Yet, this does not change the overall concept of prolonged disk lifetimes through replenishment from infall.} 
\end{appendix}
\end{document}